\documentclass[conference,a4paper]{IEEEtran}
\IEEEoverridecommandlockouts
\usepackage{cite}
\usepackage{float}
\usepackage{graphicx}
\graphicspath{{./images/}}
\usepackage[T1]{fontenc}
\usepackage{amsmath}
\usepackage{xcolor}
\usepackage{mathtools}
\usepackage{multicol}
\usepackage{multirow}
\usepackage{array} 
\usepackage{tabularx}
\usepackage{balance}
\usepackage{xcolor}
\usepackage{siunitx, booktabs}
\usepackage{ragged2e}
\usepackage{multirow}
\usepackage[ruled,vlined]{algorithm2e}
\usepackage{url}
\usepackage{soul}

\newcommand{\covid}{COVID-19}

\newcommand{\subf}[1]{#1}

\newcommand{\fref}[1]{Fig.~\ref{#1}} 
\newcommand{\sref}[1]{Section~\ref{#1}}

\newcommand{\tref}[1]{Table~\ref{#1}}
 
\newcommand{\aref}[1]{Algorithm~\ref{#1}}

\providecommand{\keywords}[1]{\textbf{\textit{Keywords---}} #1}

\usepackage[labelsep=period]{caption}

\begin{document}

\title{Spatial analysis of COVID-19 and socio-economic factors in Sri Lanka}

\author{\IEEEauthorblockN{ Rumali Perera}
\IEEEauthorblockA{\textit{Faculty of Science} \\
\textit{University of Peradeniya}\\
Sri Lanka \\
rumalip@sci.pdn.ac.lk}
\and
\IEEEauthorblockN{Harshana Weligampola}
\IEEEauthorblockA{\textit{Faculty of Engineering} \\
\textit{University of Peradeniya}\\
Sri Lanka \\
harshana.w@eng.pdn.ac.lk}
\and
\IEEEauthorblockN{Umar Marikkar}
\IEEEauthorblockA{\textit{Faculty of Engineering} \\
\textit{University of Peradeniya}\\
Sri Lanka \\
umar.m@eng.pdn.ac.lk}

\and
\IEEEauthorblockN{Suren Sritharan}
\IEEEauthorblockA{\textit{Faculty of Engineering} \\
\textit{University of Peradeniya}\\
Sri Lanka \\
suren.sri@eng.pdn.ac.lk}
\and
\IEEEauthorblockN{Roshan Godaliyadda}
\IEEEauthorblockA{\textit{Faculty of Engineering} \\
\textit{University of Peradeniya}\\
Sri Lanka \\
roshangodd@ee.pdn.ac.lk}
\and
\IEEEauthorblockN{Parakrama Ekanayake}
\IEEEauthorblockA{\textit{Faculty of Engineering} \\
\textit{University of Peradeniya}\\
Sri Lanka \\
mpb.ekanayake@ee.pdn.ac.lk}
\and
\IEEEauthorblockN{Vijitha Herath}
\IEEEauthorblockA{\textit{Faculty of Engineering} \\
\textit{University of Peradeniya}\\
Sri Lanka \\
vijitha@ee.pdn.ac.lk}
\and
\IEEEauthorblockN{Anuruddhika Rathnayake}
\IEEEauthorblockA{\textit{Faculty of Medicine} \\
\textit{University of Peradeniya}\\
Sri Lanka \\
m29782@pgim.cmb.ac.lk}
\and
\IEEEauthorblockN{Samath Dharmaratne}
\IEEEauthorblockA{\textit{Faculty of Medicine} \\
\textit{University of Peradeniya}\\
Sri Lanka \\
samath.dharmaratne@med.pdn.ac.lk}
}

\maketitle
\IEEEpubidadjcol

\begin{abstract}
The spread of the global \covid\ pandemic affected Sri Lanka similar to how it affected other countries across the globe. The Sri Lankan government took many preventive measures to suppress the pandemic spread. To aid policy makers in taking these preventive measures, we propose a novel district-wise clustering based approach. Using freely available data from the Epidemiological Department of Sri Lanka, a cluster analysis was carried out based on the \covid\ data and the demographic data of districts. K-Means clustering and spectral clustering models were the selected clustering techniques in this study. From the many district-wise socio-economic factors, population, population density, monthly expenditure and the education level were identified as the demographic variables that exhibit a high similarity with \covid\ clusters. This approach will positively impact the preventive measures suggested by the relevant policy making parties of the Sri Lankan government.
\end{abstract}



\section{Introduction}

The daily cases of the novel Coronavirus (COVID-19) are increasing rapidly throughout the world, with Sri Lanka being no exception. Along with many neighbouring countries, Sri Lanka too was affected by \covid\ with the first reported case being on the 11\textsuperscript{th} of March 2020 \cite{firstCovidInSriLanka}. The Sri Lankan government took preventive measures by enforcing lockdown strategies starting from 20th March 2020 continuing for 2 months. The continuous curfews and halting inter-district travels lead to a temporary suppression of the pandemic spread. However, studies have shown that policies and lockdown measures taken without adequate data-driven analysis results in side effects from multiple points of view\cite{pone_mental_lockdown, eci_non-evidence, pone_heartrate_lockdown}. For example, lack of social interactions has resulted in students of universities worldwide experiencing negative psychological impacts\cite{pone_mental_NJ,pone_psychological_US,pone_depression_BD,pone_edu_spain}. In an economic sense, small business owners and labour workers have been affected by excessive lockdown measures and closure of industries\cite{small_businesses}. Adverse effects have also occurred in Sri Lanka\cite{IFC2020}, mainly due to the sudden and uncontrolled lockdown procedures. This calls for optimal decision making when planning and scheduling travel restrictions and lockdown procedures.\par

Mathematically modelling \covid\ in time and space is a major contributor for data-driven analysis and optimal decision making of \covid\ related policies and lockdown measures\cite{optimal_decisions}. One such straightforward method is the use of clustering algorithms to cluster sub-regions of a larger region based on \covid\ related data. This allows for decisions to be made cluster-wise, thereby minimising economic and social effects due to lockdowns in safer regions. Clustering is the process of identifying similar points from a large sea of points and grouping them together, based on one or more properties\cite{clustering_algos}. To overcome and prevent the unfavourable situations on a community, multiple studies making use of clustering exists in literature. Clustering techniques have been used to study the hardships in living environments in India \cite{das2021living}. It was determined that clusters mainly depends on the availability of basic services for which spatial clustering has been utilized. In the \covid\ domain, a group of experts have analysed the effectiveness of dimension reduction algorithms to analyse and cluster large volumes of \covid\ genome sequences \cite{hozumi2021umap}. A spatial analysis of \covid\ in the city of New York, USA has been carried out where the effect of multiple socio-economic factors on the average number of \covid\ cases and deaths in each county is analysed\cite{cordes2020spatial}.\par

Considering Sri Lanka, a study has been carried out to predict the total number of \covid\ cases using statistical models\cite{pone_SIR_srilanka}. However, as per the authors' best knowledge, there exists a lack of published research on spatial analysis of \covid\ in Sri Lanka, and how various socio-economic factors affect the severity of \covid\ within the country. To bridge the gap in this research, the authors propose a spatial analysis of \covid\ using a clustering based model which attempts to group similar districts of Sri Lanka exhibiting similar pandemic behaviours. These groups are then compared with various socio-economic factors, which in turn will take a step towards aiding policy makers determine strategies to suppress the disease spread optimally. \par

As the case study. freely available \covid\ data from the Epidemiological Department of Sri Lanka was obtained. For this dataset, multiple pre-processing techniques were incorporated. Existing clustering algorithms were implemented to cluster these districts based on the \covid\ cases and socio-economic features. The clustering of \covid\ cases based on socio-economic features of districts was carried out to show that different demographics of different districts plays a major role in the pandemic spread. The results of this study show the correlations between a number of chosen demographic variables and \covid\ severity in district clusters of Sri Lanka. Thereby, if additional preventive measures are imposed on the districts falling in the same cluster depending on their severity, the disease spread throughout the country can be drastically reduced and the health sector of the country can plan ahead with sufficient health resources. Thus, policy makers will be capable of taking preventive measures on similar districts which will aid in suppressing the disease spread.\par

\section{Methodology}
\label{sec:meth}

\begin{figure}[htb]
    \centering
    \includegraphics[width=0.8\linewidth]{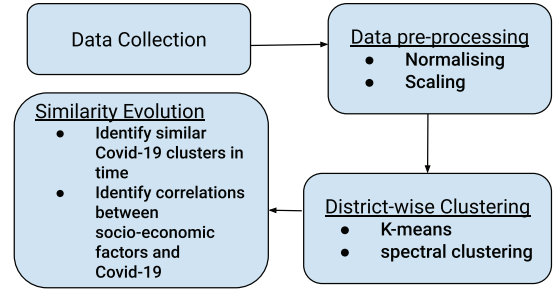}
    \caption{Process flow diagram of methodology}
    \label{fig:process_flow}
\end{figure}

The Methodology is structured as follows. First, district-wise \covid\ and relevant socio-economic and demographic data was obtained. Then, the data was pre-processed and standardised to allow for unbiased evaluations, followed by an implementation of multiple clustering algorithms to cluster districts of Sri Lanka based on \covid\ cases in time. Using this, the optimal clustering algorithm was chosen, where the cluster labels were consistent with time. This clustering algorithm was then used to cluster socio-economic variables in each district. Finally, the similarity of clusters of \covid\ and the chosen socio-economic variables were evaluated using a proposed dissimilarity metric. For further clarity, a process flow diagram of the methodology is depicted by \fref{fig:process_flow}.\par

\subsection{Data Preparation}
\label{sec:preprocess}

\begin{table*}
\caption{List of chosen demographic and socio-economic variables}
\label{tab:variables}
\centering
\resizebox{\textwidth}{!}{%
\begin{tabular}{|l|l|l|}
\hline
\multirow{2}{*}{\textbf{Feature type}} &
  \multirow{2}{*}{\textbf{Feature}} &
  \multirow{2}{*}{\textbf{Intuition}} \\
 &                                                &                                             \\ \hline
\multirow{2}{*}{Population} &
  Population &
  Potential capacity of COVID-19 spread \\ \cline{2-3} 
 & Population Density                             & Crowded-ness in public places               \\ \hline
\multirow{3}{*}{Economic} &
  Total Monthly Expenditure &
  \begin{tabular}[c]{@{}l@{}}Buying power of public translates to travelling frequency\\ to supermarkets and malls\end{tabular} \\ \cline{2-3} 
 & Monthly Expenditure on Food and non-Food items & Self-sufficency of households               \\ \cline{2-3} 
 & Poverty rate                                   & Measure of buying power                     \\ \hline
\multirow{2}{*}{Education and   Technology} &
  Persons using Internet &
  Exposure to new information regarding COVID-19 \\ \cline{2-3} 
 &
  Number of years spent in Education &
  \begin{tabular}[c]{@{}l@{}}Education level, hence the ability to process and act on\\ information regarding COVID-19 in media\end{tabular} \\ \hline
\multirow{4}{*}{Occupation} &
  Persons engaging in   skilled labour &
  \multirow{3}{*}{\begin{tabular}[c]{@{}l@{}}A measure of   social-distancing whilst carrying on their\\ professsion\end{tabular}} \\ \cline{2-2}
 & Persons engaging in   unskilled labour         &                                             \\ \cline{2-2}
 & Persons engaging in   Agriculture              &                                             \\ \cline{2-3} 
 & Unemployment rate                              & Measure of both poverty and education level \\ \hline
\end{tabular}%
}
\end{table*}

\subsubsection{Data collection}

\covid\ data was obtained from the Epidemiological Department of Sri Lanka\cite{epid_LK} . This dataset contains daily district-wise information for the total number of \covid\ confirmed cases from 15 November 2020 to 15 March 2021. Furthermore, demographic and socio-economic data was collected from the Department of Census and Statistics \cite{census_LK}  of Sri Lanka. From the large repository of data, it was important to obtain the necessary district-wise information based on how it may affect the spread and severity of \covid. Therefore, a finite number of demographic and socio-economic variables was chosen. A list of these variables and the premise at which they were chosen is shown in \tref{tab:variables}.\par

\subsubsection{Pre-processing}

The importance of data pre-processing arises in the presence of data which is badly scaled, or that contains skewed distributions. However, altering the shape of the distribution of \covid\ data would result in important information being left out of the dataset. Therefore, as a pre-processing step, multiple normalising and scaling techniques were implemented to pre-process the given dataset. Initially, the cluster analysis was carried out for the original dataset, void of any pre-processing techniques. Then, a number of techniques were implemented as summarised in \tref{tab:preprocess}.\par


\begin{table}
\caption{Summary of preprocessing techniques}
\label{tab:preprocess}

\begin{tabular}{|p{0.2\columnwidth}|p{0.45\columnwidth}|p{0.2\columnwidth}|}
\hline
\multirow{2}{*}{\textbf{Technique}} &
  \multirow{2}{*}{\textbf{Computation}} &
  \multirow{2}{*}{\textbf{Result Dataset}} \\
 &                                                &                                             \\ \hline
Population Normalisation        & Divides the number of cases by the total population of each region                           & Cases per million persons        \\ \hline
z-score standardisation         & Epicurve of a region is subtracted by its mean value, then divided by its standard deviation & Distribution centered at 0       \\ \hline
Min-max scaling (district-wise) & Values in an epicurve are divided by the maximum value of that epicurve                      & Each epicurve ranges from 0 to 1 \\ \hline
Min-max scaling (total)         & Values in an epicurve are divided by the maximum value of all epicurves dataset.             & Whole dataset ranges from 0 to 1 \\ \hline
\end{tabular}
\end{table}

\subsection{Clustering Districts of Sri Lanka Based on \covid\ and Demographic Data}
\label{sec:clustering} 

In this study, clustering techniques were carried out to cluster the districts of Sri Lanka based on \covid\ cases. In particular, k-means clustering and spectral clustering\cite{kmeans,spectral} algorithms were implemented. The choice of the aforementioned clustering algorithms was based on the principles that each algorithm uses to separate clusters.  For instance, a clustering algorithm such as Gaussian Mixture Models (GMM)\cite{GMM}, often used in image-segmentation, would not be suitable to recognise time-series data such as \covid\ cases as it is primarily a density estimation technique. Similar to clustering based on \covid, the aforementioned clustering algorithms were used to cluster districts based on socio-economic factors.\par

\subsubsection{k-means clustering}
K-means clustering is one of the basic intuitive clustering algorithms. The objective is to cluster $n$ observations in to $k$ clusters where each observation belongs to  cluster with the nearest cluster mean.\par
First, initial cluster centers are randomly initialized. Then, a cluster is assigned to each observation (point) by computing its nearest cluster center. Afterwards, a new cluster center is calculated using observations belonging to the particular cluster. This operation is repeated until clusters are converged. i.e. until the change of cluster mean after an iteration is less than a small value $\epsilon$. This algorithm results in a clustering of observations that minimizes the within cluster variance. \par

\subsubsection{Spectral clustering} 

The idea behind spectral clustering algorithm uses connectivity in a graph to cluster each node (observation) in the graph \cite{graphtheory}. The advantage of using spectral clustering is that there does not exist an assumption on the shape of the clusters.  \par
First, the observations are represented as a graph by defining adjacency matrix using a distance metric between two observations. Then, the Laplacian of the adjacency matrix is obtained. By obtaining the Eigen-values of the Laplacian, we can identify the nature of the connectivity of the graph. The first nonzero eigenvalue is called the spectral gap. The spectral gap gives some notion of the density of the graph. The second eigenvalue is called the Fiedler value, and the corresponding vector is the Fiedler vector. The Fiedler value approximates the minimum graph cut needed to separate the graph into two connected components. If the graph already has two connected components Fiedler value will be 0. Using the Fiedler vector we can identify to which connected component each node belongs.\par 
Spectral clustering algorithm looks for the first large gap between eigenvalues in order to find the number of clusters. The first eigen-vectors before this gap gives information about the cuts that will cluster the data into given number of clusters. We use K-means clustering on these first eigen-vectors to find the clustering labels of each observation.  \par

\subsection{Similarity Evaluation of Clusters}

\subsubsection{Modelling \covid\ clusters in time}
\label{sec:months}

For policy makers to make decisions based on \covid\ in a given region on a long-term basis, it is crucial that the nature of \covid\ spread exhibits similar properties in the long term. To analyse this, the dataset was first divided into four 30-day periods. Then, districts of Sri Lanka were clustered based on \covid\ cases of the first 30 days in the dataset using clustering algorithms mentioned in \sref{sec:clustering}. The change in districts in these clusters for the remaining three 30-day windows was computed according to \aref{algo:compare}. This gives a measure of the number districts belonging to a specific cluster along the time axis. In other words, the similarity that each cluster maintains throughout a longer time period is quantified.  The reason for using \aref{algo:compare} as opposed to a standard correlation analysis is due to spectral clustering being an unsupervised clustering algorithm, hence cluster labels are assigned randomly at each iteration. For example, if two clusters are generated using the same dataset at two instances, the algorithm at the first instance may label Cluster 1 as 0 and Cluster 2 as 1, whilst it may be the inverse in the second instance. Therefore, it is necessary to match the labels of clusters generated in multiple instances.\par

The optimal pre-processing technique and clustering technique among those which were mentioned previously in \sref{sec:preprocess} and \sref{sec:clustering} respectively, were chosen based on the pair of techniques which result in the highest similarity of the district clusters in time, computed using \aref{algo:compare}. In addition, the number of points in each cluster was considered, and parameters which resulted in biased clusters were avoided. This pair of techniques was then used to obtain comparisons between districts clustered according to \covid\ cases and districts clustered based on socio-economic factors.\par

\begin{algorithm}[ht]
\caption{Comparison of two sets of clustering labels}
\label{algo:compare}
\SetAlgoLined
\KwIn{Two finite sets $A=\{a_1, a_2, \ldots, a_n\}$ and $B=\{b_1, b_2, \ldots, b_n\}$  of integers representing two sets of labels. Number of regions $n$.}
\KwOut{Dissimilarity between two sets of clustering labels}
 $best \gets \inf$\;
 $P \gets permutate([1, 2, \ldots, n])$
 \For{each $p \in P$}{
    
  $\hat{A} \gets $ remapped $A$ according to $p$\;
  $cost \gets \frac{\sum_{i\in[1\ldots n]}(\hat{A}_i-A_i)^2}{n}$\;
  \If{$cost < best$}{
   $best \gets cost$\;
   }
 }
\end{algorithm}

\subsubsection{Comparing effects of socio-economic factors on \covid\ clusters}

To analyse the effect of socio-economic factors on how districts have been clustered based on \covid\ cases, a similar approach to \sref{sec:months} was carried out. Districts of Sri Lanka were clustered based on each socio-economic variable mentioned in \tref{tab:variables}. As each variable is one-dimensional and each district corresponds to a single scalar value, cluster indexes were pre-set in increasing order. This method of clustering provides meaning and context to the output clusters. For instance, districts which belong to the cluster based on 'population' with the smallest centroid value, will contain the districts with the smallest populations. The dissimilarity metric ($SM1$) between these clustered districts based on each socio-economic variable and \covid\ were computed using \aref{algo:compare}, similar to \sref{sec:months}.\par

To outline the differences between the dissimilarity metrics ($SM1$) of two similar cluster outputs vs dissimilar cluster outputs, \aref{algo:compare} was carried out for both cluster outputs being known (which produces $SM1$), and also with one of the cluster outputs being a randomised array including cluster labels. For example, if district clusters based on population and \covid\ were to be compared, a dissimilarity metric would also be obtained for district clusters obtained population and a randomly generated array of cluster labels. A Monte Carlo simulation would be carried out to obtain the mean dissimilarity ($SM2$) between the known cluster output and the randomised array. In this study, a total of 100 Monte Carlo simulations was carried out for such an instance. If $SM1<<SM2$, it would imply that there is a high similarity between the two chosen cluster outputs. \par

All code has been written in Python 3.8, on the online Google Collaboratory software (colab.google.io). Tensorflow 2.0 has been used as a Machine Learning tool, in addition to the conventional data science libraries in Python 3.8. The complete codebase can be found at: \url{https://github.com/pdncovid/covid_clustering}.\par


\section{Results and Discussion}

\begin{figure*}[htb]
\centering
\begin{tabular}{c c c}
    (i) &
    \subf{\includegraphics[width=60mm,trim=5mm 0cm 0cm 0cm, clip=true]{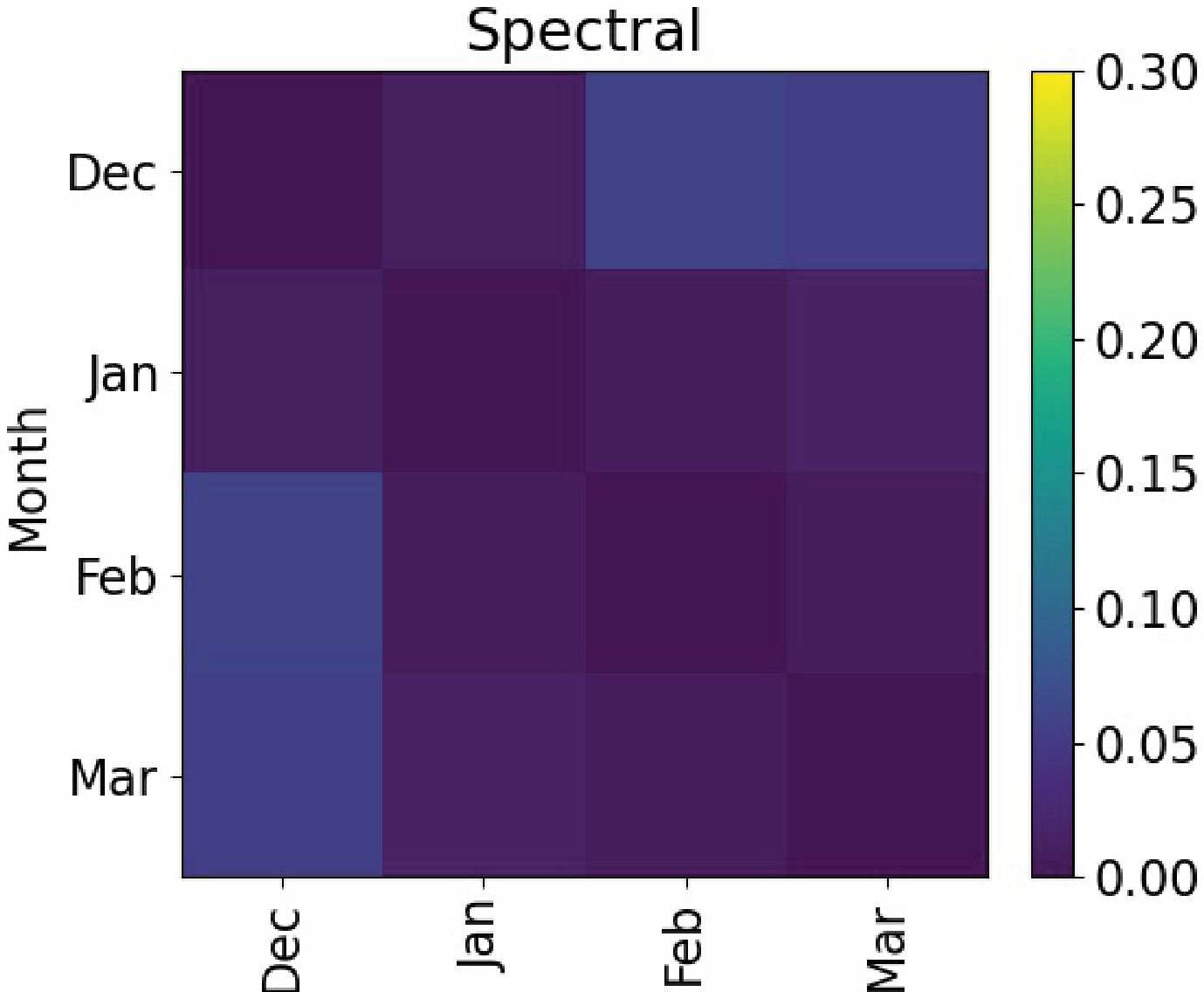}} &
    \subf{\includegraphics[width=60mm,trim=5mm 0cm 0cm 0cm, clip=true]{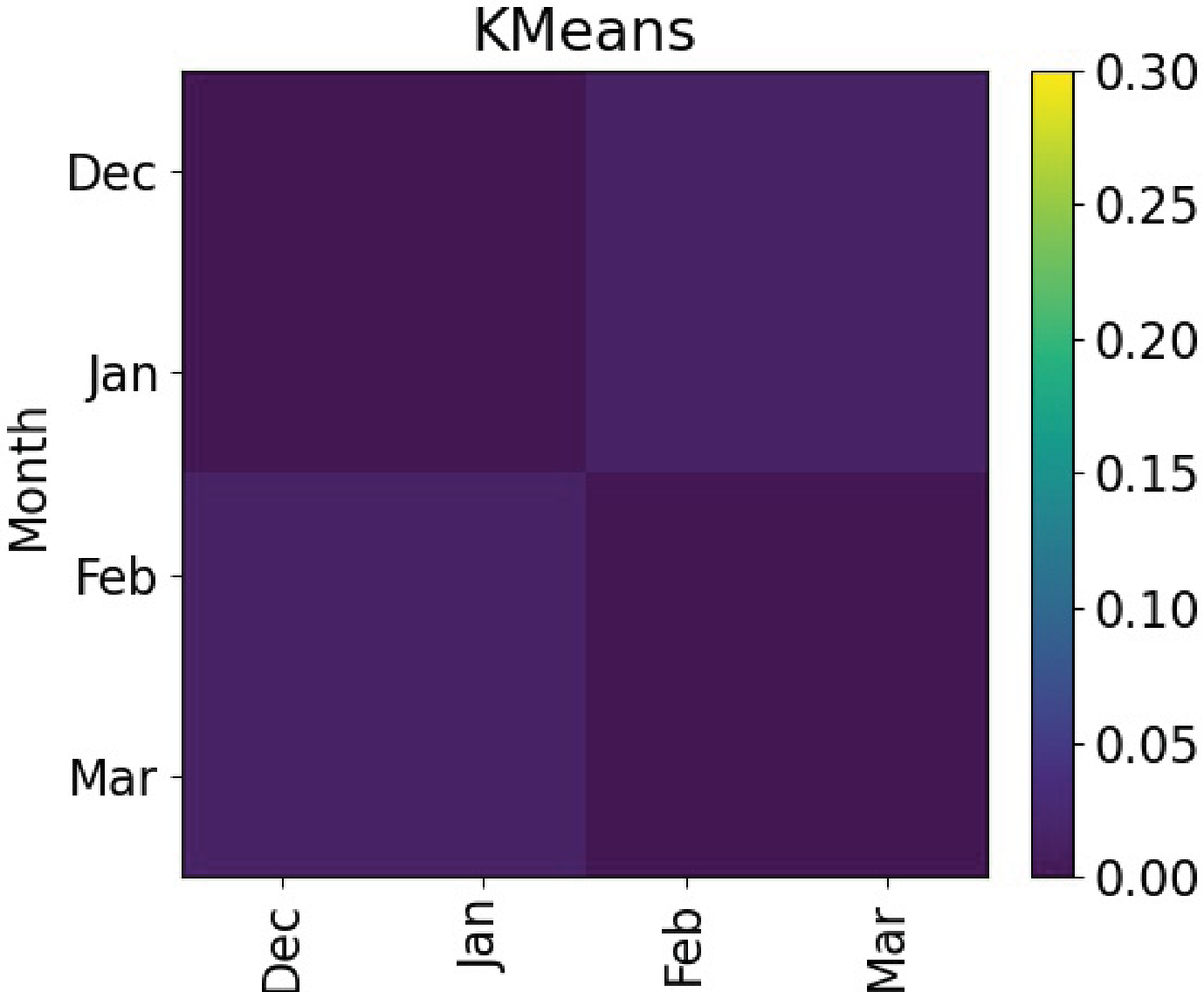}} 
    \\[0pt]
    (ii) &
    \subf{\includegraphics[width=60mm,trim=5mm 0cm 0cm 0cm, clip=true]{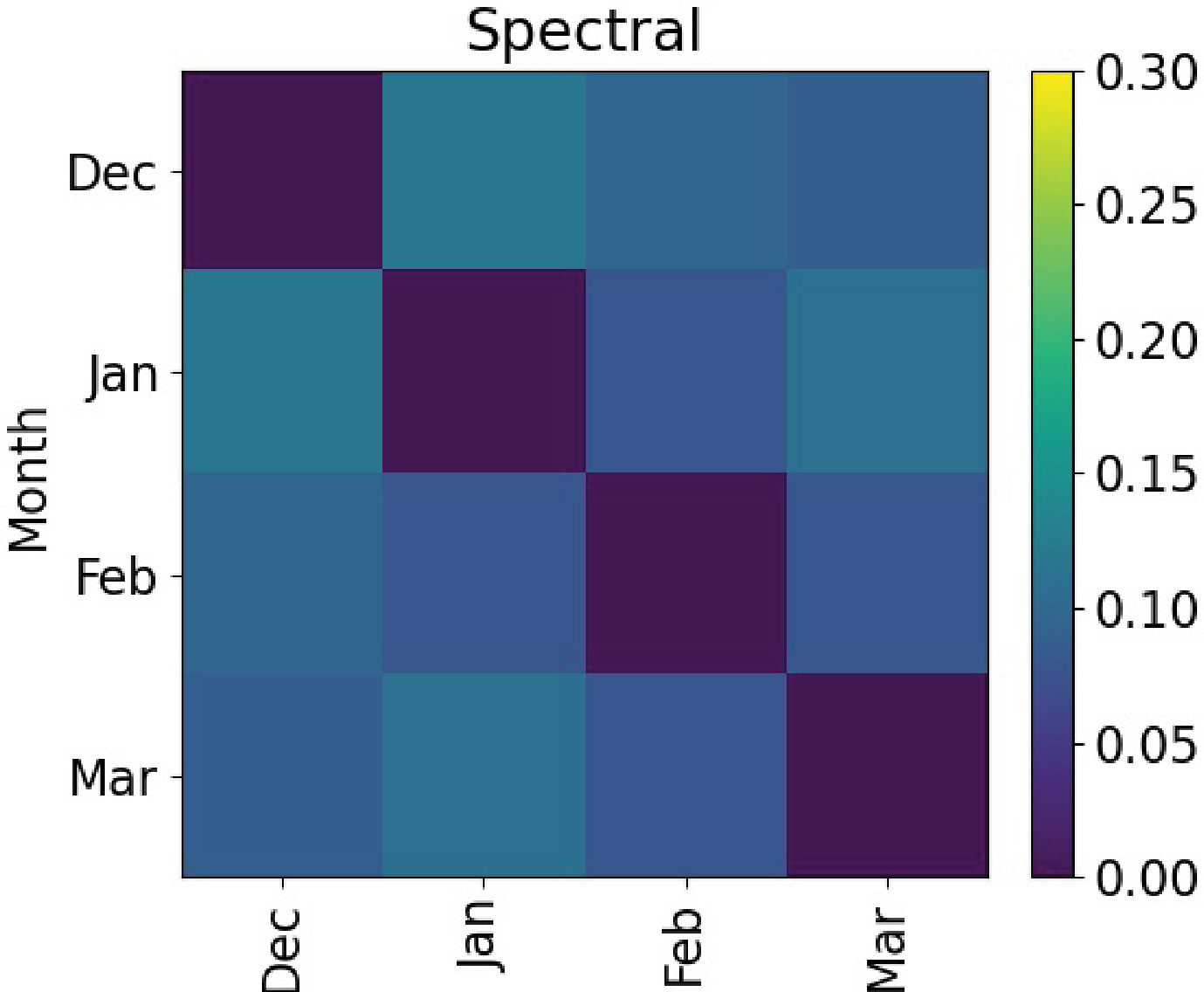}} &
    \subf{\includegraphics[width=60mm,trim=5mm 0cm 0cm 0cm, clip=true]{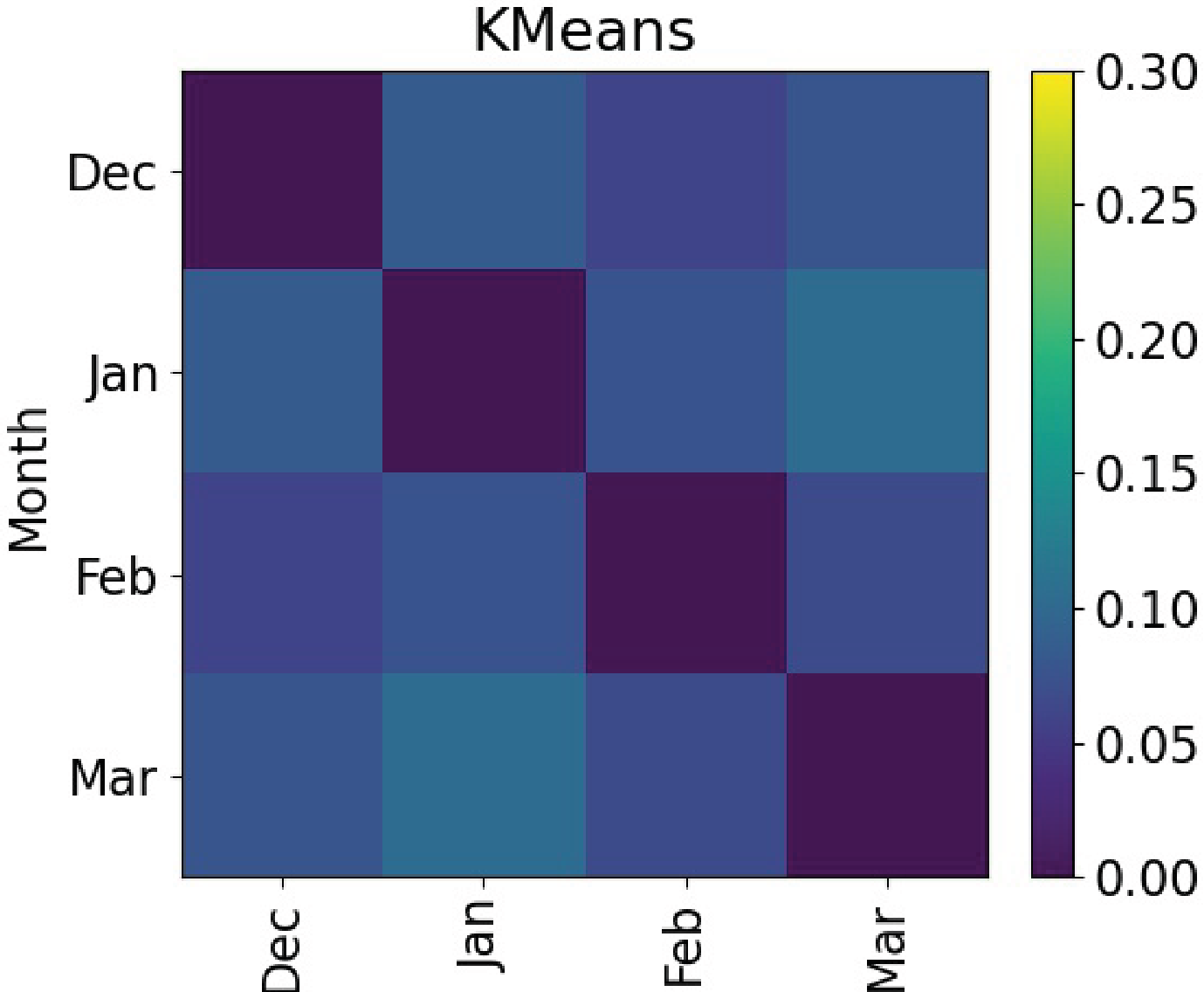}} 
    \\[0pt]
     &
    (a) &
    (b) 
\end{tabular}
\caption{Comparison of dissimilarity metric of \covid\ clusters between each month. Row (i) uses the original (non-scaled) data and row (ii) uses min-max normalised data. Column (a) uses Spectral Clustering algorithm and column (b) uses K-means clustering algorithm. Lower values (darker blue) correspond to a high similarity (low dissimilarity) between two clusters, whereas high (blue-green) depicts a lesser correlation between two clusters.}
\label{fig:month_vs_month}
\end{figure*}

\subsection{District-wise Clusters Based on \covid\ Cases}

\fref{fig:month_vs_month} shows the dissimilarity metric of districts clustered using k-means clustering based on \covid\ cases in each month, for both original and min-max normalised data. In this figure, the lighter shade (higher end of the spectrum) corresponds to higher dissimilarities. It can be observed that both spectral clustering and k-means clustering show high similarity between months, for original data over normalised data. This is due to the mean value of each epicurve contributing to the clustering process. Similar to \fref{fig:month_vs_month}, comparisons were carried out for the remaining normalisation techniques in \tref{tab:preprocess}, and original data was chosen as it contained the highest similarities between months.\par

\raggedbottom

Although k-means clustering exhibits a higher similarity between months when compared to spectral clustering even on original data, upon analysis of the number of points in each cluster, it was observed that almost all the points belong to a single cluster. This resulted in a biased cluster output, hence an omission of the k-means algorithm in the evaluations between clusters of \covid\ and socio-economic variables. \par

\begin{figure}[htb]
    \centering
    \includegraphics[width=\linewidth]{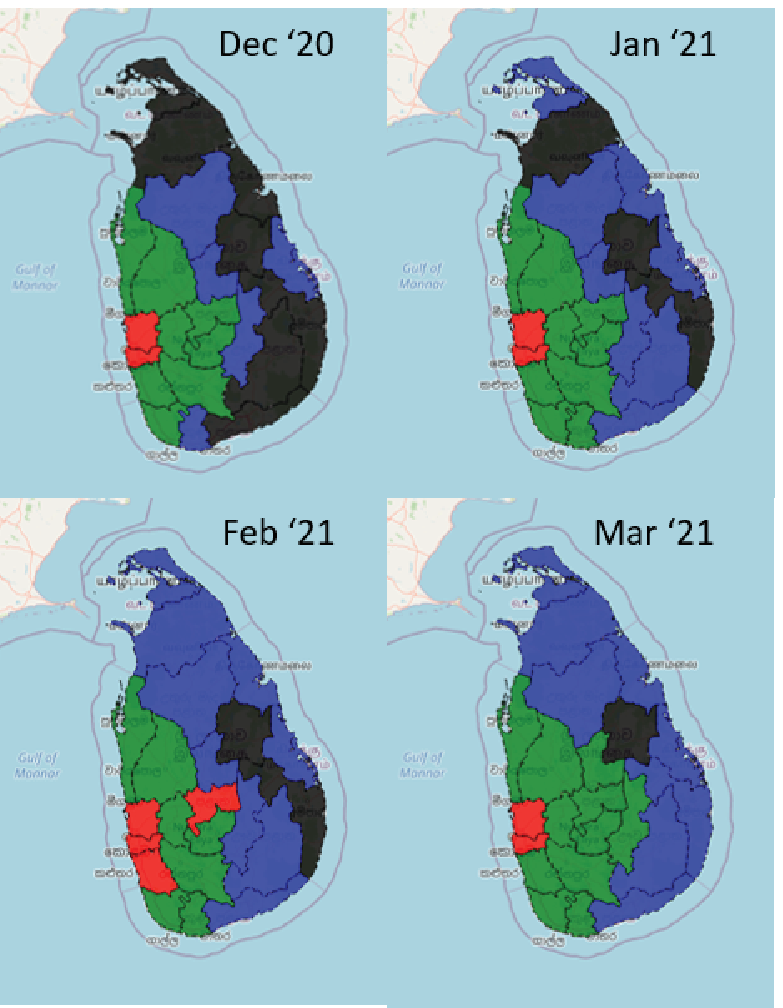}
    \caption{Clustering districts using \covid\ cases in each month. Each cluster label (colour) is assigned as mentioned in Algorithm 1.}
    \label{fig:month_maps}
\end{figure}

In contrast to k-means clustering, spectral clustering algorithm tends to optimally balance out the number of points in each output cluster. Therefore, spectral clustering using original \covid\ data was used for further evaluations. A map of clustered districts in each month is shown in \fref{fig:month_maps}. Here, it is observed that the districts belonging to each cluster change minimally when moving from one month to another. This allows for the assumption that the nature of \covid\ in a given region remains unchanged roughly throughout a period of one month, hence allowing for mid to long-term planning of lockdowns and other preventive measures.\par

\subsection{Similarity Between \covid\ Clusters and Socio-economic Variables}

\begin{figure}[H]
    \centering
    \includegraphics[width=\linewidth]{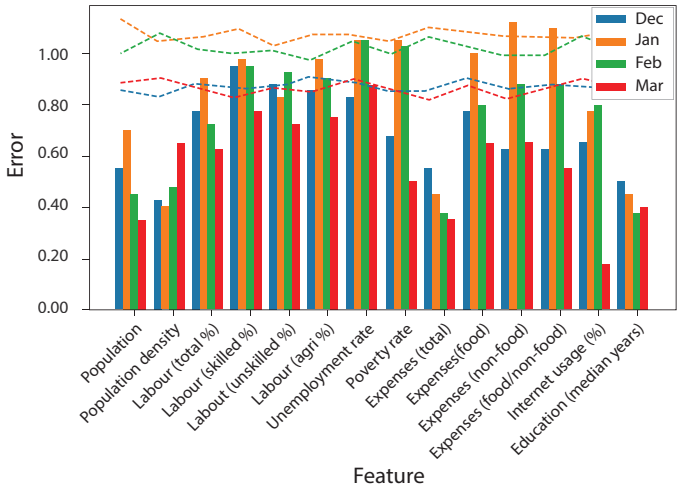}
    \caption{Comparing dissimilarity of clustering using features with clustering using \covid\ cases. Dashed lines represents dissimilarity when features are clustered using random labels. A bar plot significantly below the dashed line represents an existence of correlation between \covid\ clusters and feature clusters.}
    \label{fig:covid_vs_demo}
\end{figure}

\begin{figure}[H]
    \centering
    \includegraphics[width=\linewidth]{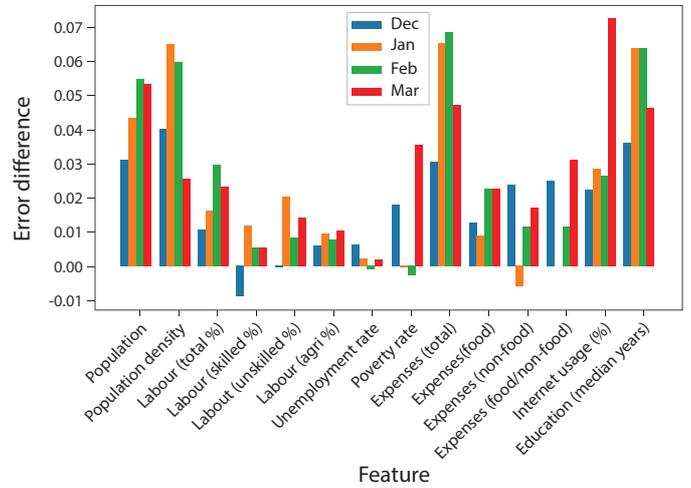}
    \caption{Deviation of dissimilarity metric computed from \covid\ and computed feature clusters vs. \covid\ and random feature clusters. An inverse representation of \fref{fig:covid_vs_demo}. Higher values correspond to a high correlation between \covid\ and a given feature.}
    \label{fig:covid_vs_demo2}
\end{figure}

\fref{fig:covid_vs_demo} depicts the dissimilarity metrics between each socio-economic variable and \covid\ cluster for each month. To better represent the effect of socio-economic variables on \covid\, the difference between random cluster and known cluster dissimilarities is computed for each socio-economic variable, as shown in \fref{fig:covid_vs_demo2}. It can be observed that Population, Population density, Total monthly expenses and Median years spent in education have a low dissimilarity metric, as opposed to comparisons with other features and random labels. Another important observation is that the low dissimilarity of the these factors is consistent throughout the four months of study. This implies that policy makers should take into consideration the aforementioned socio-economic factors in their decision making process. \par

\section{Conclusion and Future Directions}

This study analyses the effect of socio-economic factors on \covid\ in Sri Lanka. The authors propose spectral clustering on non-normalised data as the optimal technique to cluster districts of Sri Lanka based on \covid. Out of numerous socio-economic variables, population, population density, monthly expenditure and education level are suggested as the main factors that policy makers should consider when enforcing policies such as lockdown measures and travel restrictions. As the districts clustered based on \covid\ cases contain high similarity throughout the time period, the proposed clustering technique will also promote localised decision making to control \covid\ in the country.\par 

This approach of data-driven analysis aids to bridge the gap between optimal and non-optimal decision making for \covid\ policies in Sri Lanka. A challenge encountered during this research was the presumed disparity between actual \covid\ cases and observed \covid\ cases, thereby depicting an inaccurate measure of \covid\ severity. The authors aim to analyze the actual infected cases from the observed tested cases, using an AI-based simulation model. This will further alleviate the quality of \covid\ research in Sri Lanka, as it can be utilized as a foundation for many other epidemiological models and compartmental models which will aid in a better understanding of the pandemic situation. \par

\bibliographystyle{IEEEtran}
\bibliography{main}
\clearpage

\end{document}